\documentclass{article}
\usepackage[margin=2cm]{geometry}  
\usepackage{url}

\usepackage[
    backend=biber,
    style=numeric,
    citestyle=numeric,
    doi=false,
    isbn=false,
    url=false,
    maxnames=50,
    firstinits=true]{biblatex}
\usepackage{authblk}
\usepackage{physics}

\usepackage{graphicx} 
\usepackage[hidelinks]{hyperref}
\usepackage{xcolor}
\usepackage{caption}
\usepackage{subcaption}
\usepackage{amsmath, amssymb, amsthm}
\usepackage{mathtools} 
\usepackage{soul}
\usepackage{siunitx} 
\usepackage{enumerate}
\usepackage{enumitem}
\usepackage{float} 
\usepackage{tikz} 
\usetikzlibrary{matrix}

\theoremstyle{definition}

\title{On the impact of geometric variance on the performance of formed parts: 
A probabilistic approach on the example of airbag pressure bins}
\date{}
\author[1]{Lukas~Schnelle}

\affil[1]{Institute of Structural Mechanics and Lightweight Design, RWTH Aachen University, W\"ullnerstr. 7, 52062 Aachen, Germany}

\author[2]{Niklas Fehlemann}
\affil[2]{Institute of Metal Forming, RWTH Aachen University, Intzestr. 10, 52072 Aachen, Germany}

\author[3]{Ali O.M.~Kilicsoy}
\affil[3]{Chair for Reliability Engineering, TU Dortmund University, Leonhard-Euler-Strasse 5, 44227 Dortmund, Germany}

\author[4]{Niklas Bechler}
\affil[4]{Institute of Forming Technology and Lightweight Components, TU Dortmund University, Baroper Str. 303, 44227 Dortmund, Germany}

\author[3]{Marcos A.~Valdebenito}

\author[4]{Yannis P.~Korkolis}

\author[3]{Matthias G.R.~Faes}

\author[2]{Sebastian M\"unstermann}

\author[1]{Kai-Uwe~Schr\"oder}

\begin{document}

\maketitle

\begin{abstract}

Scatter in properties resulting from manufacturing is a great challenge in lightweight design, requiring consideration of not only the average mechanical performance but also the variance which is done e.g., by conservative safety factors. One contributor to this variance is the inherent geometric variability in the formed part. To isolate and quantify this effect, we present a probabilistic numerical study, aiming to assess the impact of geometric variance on the resulting part performance. By modelling geometric deviations stochastically, we aim to establish a correlation between the variance in geometry with the resulting variance in performance. The study is done on the example of an airbag pressure bin, where a better understanding of this correlation is crucial, as it allows for the design of a lighter part without changing the manufacturing process. Instead, we aim to implement more targeted and effective quality assurance, informed by the performance impact of geometric deviations.

\end{abstract}

\section{Introduction}\label{sec:intro}
One central goal of lightweight design is to fully use the structure that is given. This requires in particular, that the expected loads do not exceed the loads a structure can withstand, times a given safety factor. However, if the difference between these values is too big, there is potential for a lighter structure that can fulfil the same purpose. The load-bearing capacity of a structure is typically determined through experimental testing on a number of samples. These inevitably will have some scatter in them, with one major source being manufacturing variances. In metal forming these variances often manifest as geometric variations from the target geometry \cite{manque2025, hering2020, gitschel2022, tekkaya2017}.
To accommodate these variances in the geometry and therefore performances, often conservative safety factors are used, so e.g., the load that is calculated with is the $99$-th percentile of the available tests, so only $1\%$ of the available data is below the value used. The amount of variance correlates to the potential for a lighter part, as with a large variance in the scatter the acceptable load is reduced and thus the part is made heavier. Therefore, one approach is to reduce this variance as much as possible \cite{gitschel2025}.
In literature there are some discussions of the influence of the geometric variance stemming from the forming process, however most of them focus on sheet metal forming, with a focus on the successful forming process \cite{radi2007, kleiber2002}. These models seem to rely heavily on the fact, that the formed part starts as a sheet. In \cite{soderberg2017} the authors discuss the creation of a digital twin, with a particular focus on archiving geometries which are as accurate as possible, while the performance impact of a geometric variation is used as a motivation, but the resulting mechanical performance not used in the presented model. Moreover, the authors of \cite{bohm2022} discuss the influence of geometric imperfections under internal negative pressure, with particular focus on the buckling problem. 
In this work, we pursue to address the influence which the geometric variance, stemming from the manufacturing process, has on the scatter which can be observed in the performance. Particularly we focus on the example of an airbag pressure bin, produced by cold forging an elongated part with an internal cavity. Rather than focusing solely on reducing this variance in the scatter, we aim to identify and quantify its sources by correlating geometric variations with the resulting scatter in mechanical performance. To this end, we employ a probabilistic framework, using variance-based sensitivity analysis, to achieve this correlation on the example of airbag pressure bins on artificially varied geometries. This method, once validated, allows for a more accurate performance prediction for an individual component post-manufacturing, after its specific geometry has been measured.
By integrating this predictive capability into quality assurance, we enable the ability to produce lighter parts which are verifiably able to withstand the required loads in the future.

\section{Stochastic sensitivity analysis}\label{sec:stochastics}

In this section, we seek to elaborate the analysis of the variance-based sensitivity of a stress state with regards to the underlying geometry, which will be discussed in detail in the following section. Note, that while we are interested in the scatter, it is only qualitative and thus we define a quantitative measure which is mathematically defined in order to evaluate the scatter. In particular applying probability theory we assume a variance of the geometry by a probability distribution. The stresses within the geometry are defined as scalar quantities, indexed spatially in three dimensions per \(\mathbf{X},\mathbf{Y},\mathbf{Z}\) in the FE analysis. These responses depend on the chosen geometry parameters \emph{angle of attack} \(\phi\) and \emph{additional indentation} \(h\), which will be described in the following section, and now collectively defined as a parameter vector \(\mathbf{\theta}\). We assume these parameters to be uncertain and quantify their uncertainty and its propagation via probability theory \cite{elnashai1991}. We then define vector \(\mathbf{\theta}\) as a single realization of a random variable vector \(\mathbf{\Theta}\). In general, through prior knowledge and data, probability density functions \(f_{\Theta_{i}}(\theta_{i})\) for \(i = 1,\ldots,n_{\mathbf{\theta}}\) are selected; with \(n_{\mathbf{\theta}} = 2\) for this case. Uncertainty associated with both geometrical parameters is described considering uniform distributions. Regarding the behavior of airbag pressure bins, we consider the response \(r\) at any space of the geometry as the maximum von Mises stress \(\varsigma\) as our quantity of interest. Under a given set of values of the geometry parameter vector \(\mathbf{\theta}\) along the aforementioned boundary constraints we define

\begin{equation}
r = \max_{\left( \mathbf{X},\mathbf{Y},\mathbf{Z} \right)\mathbb{\in G}\left( \mathbf{\theta} \right)}\varsigma\left( \mathbf{X},\mathbf{Y},\mathbf{Z;\theta} \right) = g\left( \mathbf{\theta} \right)
\label{eq:one}
\end{equation}

\(\mathbb{G(}\mathbf{\theta)}\) denoting the geometry domain as it is defined by \(\mathbf{\theta}\). As the maximum stress is uncertain, due to uncertain geometry parameters in \(\mathbf{\theta}\), with \(r\) being a single realization of the random variable \(R,\) whereby \(R\ \)is sampled in our case from the FE analysis model. This defines the FE analysis process as a black box and allows the quantification of uncertainty per mean and variance with variance-based sensitivity analysis. The first and second order statistics of \(r\) can be computed as

\begin{equation}
\mathbb{E}_{\mathbf{\Theta}}(r) = \int_{\mathbf{\theta} \in \Omega_{\mathbf{\theta}}}^{}g\left( \mathbf{\theta} \right)f_{\mathbf{\Theta}}\left( \mathbf{\theta} \right)d\mathbf{\theta}
\label{eq:two}
\end{equation}

\begin{equation}
\mathbb{V}_{\mathbf{\Theta}}(r) = \int_{\mathbf{\theta} \in \Omega_{\mathbf{\theta}}}^{}\left( g\left( \mathbf{\theta} \right) - \mathbb{E}_{\mathbf{\Theta}}(r) \right)^{2}f_{\mathbf{\Theta}}\left( \mathbf{\theta} \right)d\mathbf{\theta,}
\label{eq:three}    
\end{equation}

with \(\Omega_{\mathbf{\theta}}\) denoting the support of \(\mathbf{\theta}\), expected value \(\mathbb{E}_{\mathbf{\Theta}}\) and variance \(\mathbb{V}_{\mathbf{\Theta}}\). Subsequent variance-based global sensitivity analysis can occur with these properties.

In order to evaluate the effect of the chosen geometry parameters (components of~\(\mathbf{\theta}\)) on the variance of the quantity of interest, we apply sensitivity analysis via Sobol' indices \cite{sobol1993}. Sobol' indices are bounded from \(0\) to \(1\), with \(1\) denoting the highest impact. For first-order Sobol' sensitivity indices, \(S_{i}\), we have

\begin{equation}
S_{i} = \frac{\mathbb{V}_{\Theta_{i}}\left( \mathbb{E}_{\mathbf{\Theta}_{- i}}\left( r|\theta_{i} \right) \right)}{\mathbb{V}_{\mathbf{\Theta}}(r)}\ for\ i = 1,\ldots,n_{\mathbf{\theta}},
\label{eq:four}
\end{equation}

with \(\mathbb{V}_{\Theta_{i}}\) denoting the variance with respect to only \(\Theta_{i}\). \(\mathbf{\Theta}_{- i}\) denotes the vector containing all random variables of geometry parameters, except for \(\Theta_{i}\). This isolates the effect of \(\Theta_{i}\) on the variance in relation to the total variance, thus the first-order Sobol' index represents the main effect of a parameter on the variance of the response of interest without considering interaction with other
parameters. This also means, that the sum of first-order Sobol' indices indicate the share of main effects of the chosen parameters on the variance of the response of interest without interactions. Generally, we can also capture the total effect of a geometry parameter on the variance of a response of interest, that is, the main effect and also the influence due interaction with other parameters, as per the total Sobol' sensitivity indices \(S_{T_{i}}\) \cite{saltelli2008}. We define \(S_{T_{i}}\) as

\begin{equation}
S_{T_{i}} = \frac{\mathbb{V}_{\mathbf{\Theta}}(r) - \mathbb{V}_{\mathbf{\Theta}_{- i}}\left( \mathbb{E}_{\Theta_{i}}\left( r|\mathbf{\theta}_{- i} \right) \right)}{\mathbb{V}_{\mathbf{\Theta}}(r)}\ for\ i = 1,\ldots,n_{\mathbf{\theta}},
\label{eq:five}
\end{equation}

which is equal or larger than the first-order Sobol` indices as per its definition. Before computing the Sobol' indices, we apply a surrogate model, in this case a Gaussian process regression (GPR), which provides simple evaluation of first and second order statistics by design \cite{rasmussen2006}. This enables us to avoid repeated computationally expensive FE analyses. The GPR involves setting up a prior via mean \(\mu_{I}\), the expected value of \(r\) at \(\mathbf{\theta}\) and a covariance function \(\kappa_{I}\), capturing joint variability for two separate realizations \(\mathbf{\theta}\) and \(\mathbf{\theta}'\) of the geometry parameters, between \(r\) and \(r'\). We set \(\mu_{I} = \mu_{0}\) as a constant, and \(\kappa_{I}\) as a squared exponential

\begin{equation}
\kappa_{I}\left( \mathbf{\theta},\mathbf{\theta}' \right) = \sigma_{0}^{2}\exp\left( \frac{1}{2}\sum_{i = 1}^{n_{\mathbf{\theta}}}\frac{\theta_{i} - {\theta'}_{i}}{L_{i}^{2}} \right),
\label{eq:six}
\end{equation}

where \(\sigma_{0}^{2}\) represents the variance of the prior, \(L_{i}\) the respective scale length attributed to geometry parameters \(\theta_{i}\). The small sample data, \((\mathbf{r}_{d},\mathbf{\Theta}_{d})\), generated by the FE analysis, is used to fit the GPR, with \(\mathbf{r}_{d}\) being a vector of dimension \(n_{d}\) and \(\mathbf{\Theta}_{d}\) a matrix of dimension \(n_{d} \times n_{\mathbf{\theta}}\). The samples are generated using a Latin hypercube sampling design \cite{mckay1979} involving a total of $100$ realizations. The hyperparameters of the GPR are determined through maximum likelihood estimation on this sample. This gives us a posterior mean \(\mu_{p}\) and a posterior covariance \(\kappa_{p}\), capturing the intrinsic system response due geometry parameters, which the FE analysis evaluates. This allows cheap and efficient computation of said system's responses of interest \(\mu_{P}(\mathbf{\theta})\) in unsampled regions of \(\mathbf{\theta}\), but also quantifies the degree of uncertainty of the generated responses as per \(\kappa_{P}(\mathbf{\theta},\mathbf{\theta}').\) 

Subsequently, probabilistic descriptors of the GPR are computed. This includes the expectation's mean and variance, as well as of the variance itself. With these, the Sobol' indices can be computed. For the detailed derivations and specific expressions in order to compute these probabilistic descriptors, we direct to \cite{song2022}. Note that we require and assume standard Gaussian samples and our data was sampled uniformly. As such, a mapping per inverse CDF to standard Gaussian was done. Geometry parameters \(\mathbf{\theta}\) are normalized and response of interest \(r\) is standardized before fitting the GPR.

The resulting Sobol' indices of this analysis are discussed in Section
\ref{sec:results}.

\section{Methodology}\label{sec:methods}
\begin{figure}
    \centering
    \includegraphics[width=0.5\textwidth]{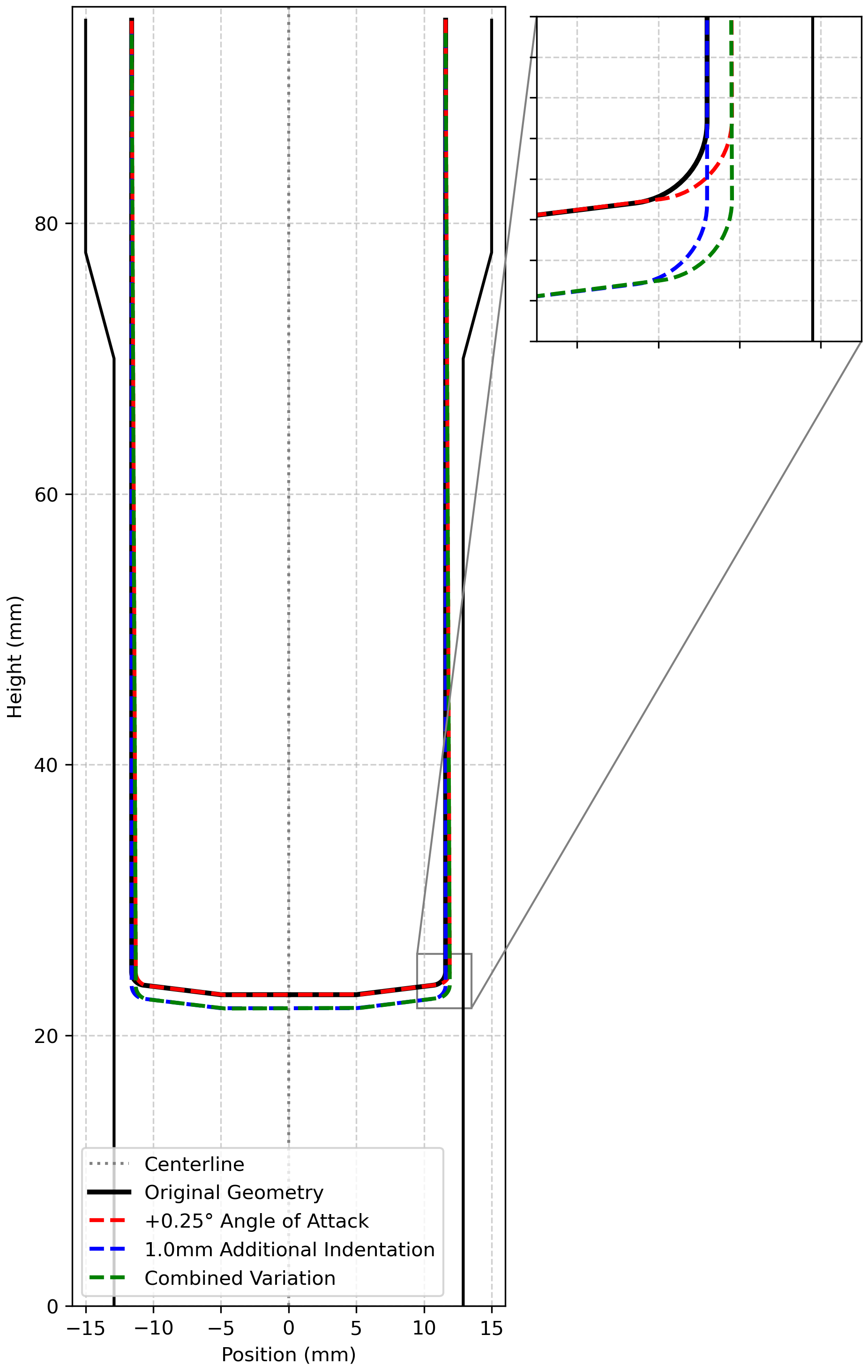}
    \caption{Schematic view of the considered geometry and variation.}
    \label{fig:geom}
\end{figure}

Our work is concerned with an airbag pressure bin, produced by cold forging 16MnCrS5 steel. This forging produces an elongated part with a cavity, in which a small pyrotechnical charge will be set off when the airbag is deployed. Thus, our load case is a one-time pressure load on the interior walls. The internal cavity with a diameter of $23.2~\mathrm{mm}$ is produced by backward can extrusion followed by a forward hollow extrusion. For the following we assume that the geometry \(\mathbb{G}\) resulting this forming process is given by the parameter vector \(\mathbf{\theta},\) which consists of two uncertain geometry parameters: firstly, the angle with which the punch was inserted into the sample during the backwards can extrusion process, later referred to as the \emph{angle of attack} \(\phi\). The second parameter is the amount of additional indentation that the subtracted cylinder has applied to it, which will be referred to as \emph{additional indentation} \(h\). These parameters were chosen to be conservative anticipations of variations in the forming process and can be seen in Figure \ref{fig:geom}. The amounts were chosen to be \(0 - {0.25}^{\circ}\) for the angle of attack and \(0\ \mathrm{mm} - 1\ \mathrm{mm}\) for the additional indentation. These values result in a geometry which has a minimum wall thickness of just under \(1\ \mathrm{mm}\), which is a deviation of \(\approx 0.3\ \mathrm{mm}\) from the target wall thickness.

The modelling approach chosen in this work was to establish a probabilistic simulation workflow. The foundation is a python script, which generates geometries with the described variations applied to it.

Next, \(100\) pairs of uniformly distributed values of the angle of attack and additional indentation,\(\phi\) and \(h,\) over their corresponding ranges were generated using Latin hypercube sampling \cite{mckay1979}. These random variables represent the variance we assume to be present in the samples due to the manufacturing process. The values were then used to generate the corresponding \(100\) geometries.

After the geometries are generated and a FE analysis is performed, using Abaqus/standard \cite{dassault2024}. For these defined geometries \(\mathbb{G(}\mathbf{\theta)}\) the inner cylinder is loaded with a pressure of \(50\ \mathrm{MPa}\). Additionally, the bottom is fixed as a boundary condition to prevent the whole geometry from moving. The simulation is run for \(4\ \mathrm{ms}\) and the response \(r\), which is the maximum of the von Mises stresses \(\varsigma\) as per Eq. \ref{eq:one}, are extracted from the simulation. This choice is made with the assumption that for the burst failure the maximum stress will dictate the failure initiation and thus correlates to the ultimate load. These values are collected in a CSV table, which is also included with the data for this work. The material model chosen is that of a cold forging steel, namely 16MnCrS5. The Young's modulus is \(207\ \mathrm{GPa}\) with a Poisson's ratio of \(0.3\). Further, \cite{hering2019} provides a flow curve which is applied to the material adjusted for a strain rate of \(1.3\). This material model does not account for properties resulting from the forming process, such as work hardening, residual stresses, etc.

As a last step we apply the sensitivity analysis as described in Section \ref{sec:stochastics} and compute the Sobol' indices of first and total order.

\begin{figure}
    \centering
    \includegraphics[width=0.5\textwidth]{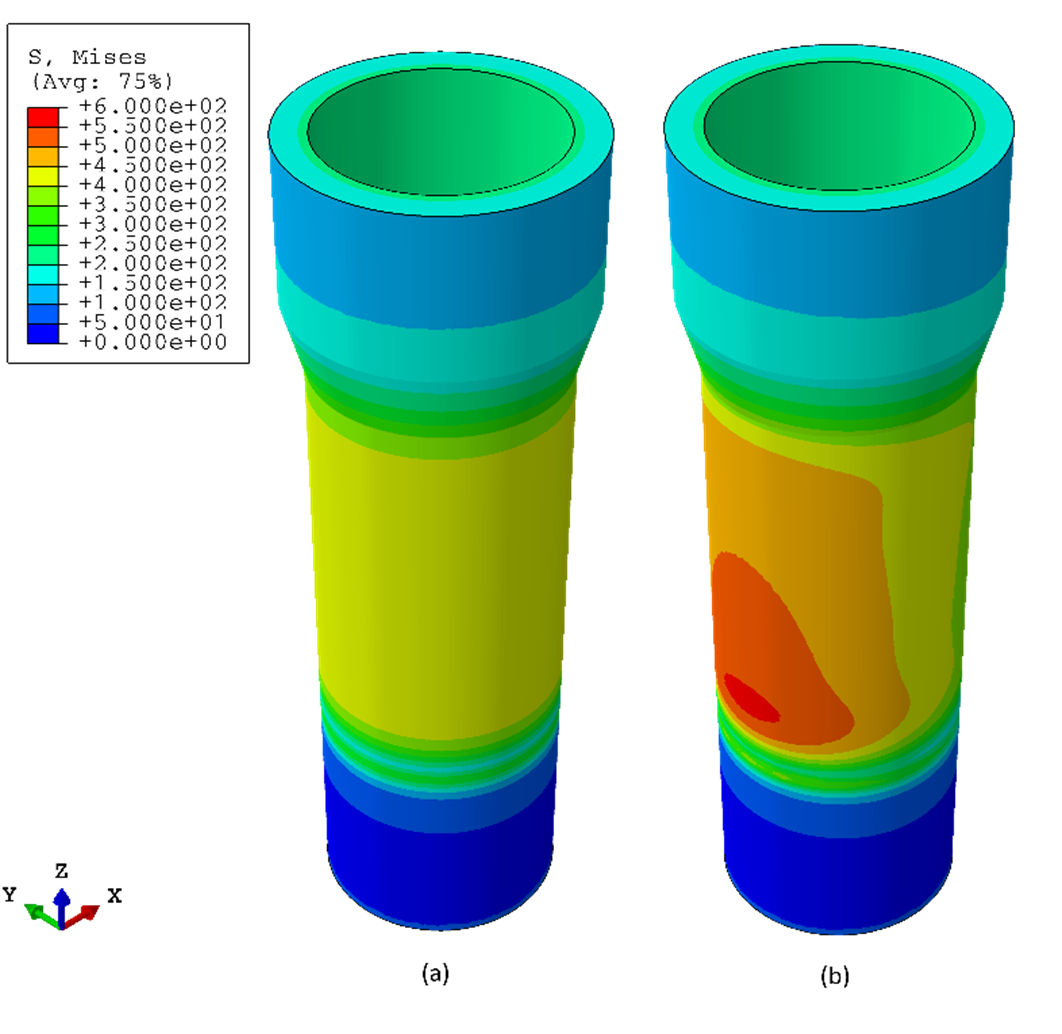}
    \caption{FEM analysis; nominal geom. (a), perturbed geom. (b, sample 31).}
    \label{fig:bins}
\end{figure}

Two example results from the FE analysis are shown in Figure \ref{fig:bins}.

\begin{figure}[H]
    \centering
    \includegraphics[width=0.5\textwidth]{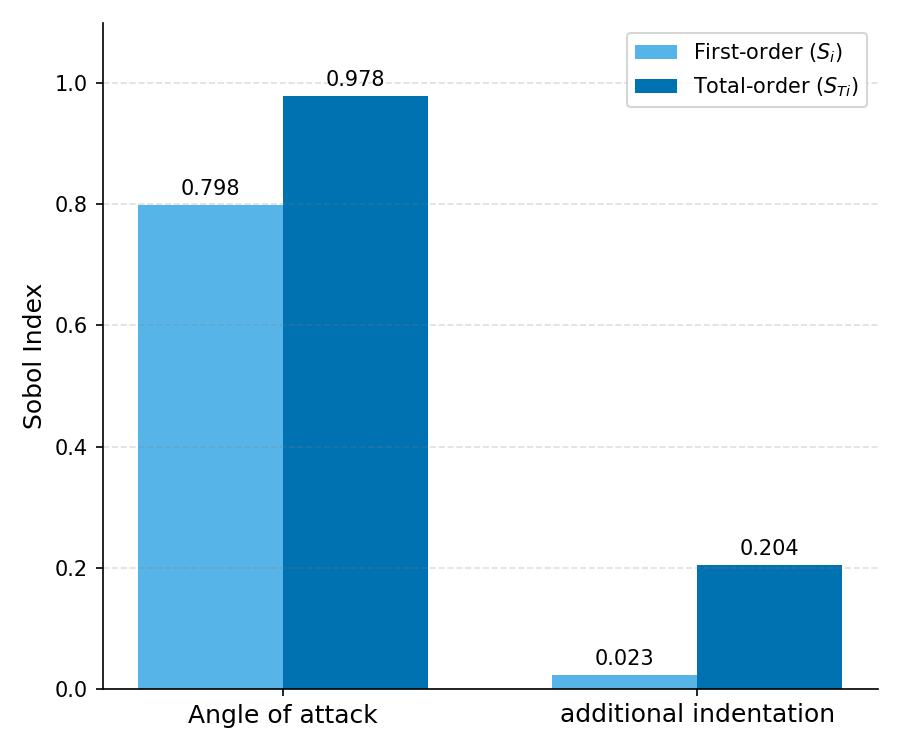}
    \caption{Sobol' indices resulting from the variations.}
    \label{fig:sobol}
\end{figure}

\section{Results and discussion}\label{sec:results}
As previously mentioned, the geometry parameters were defined such that the smallest wall thickness would be around \(\sim 1\ \mathrm{mm}\). Generally, comparing best case to worst case geometry, the variation of performance was \(\sim + 20\%\), which is a big indicator that the traditional approach of conservative safety factors would have a significant impact on the weight of the part. Even before sensitivity analysis, the samples show considerable increase in stress for the worst-case geometry.

The sensitivity analysis shows, that very minor variations in the angle of attack with which the punch is applied can have severe influence on the performance of the formed part. The total-order Sobol' indices show that the interaction effect of these two parameters is not as significant. The additional indentation does not seem to affect the performance much as its first-order Sobol' index is quite small, especially contrasted to that of the angle of attack. Its impact is only exacerbated when combined with a larger angle of attack but then it is still quite negligible. The impact of the angle of attack to the considered parameter is most significant. Its first-order Sobol' index indicates the angle to be the crucial parameter.

These results allow a better understanding of the sources of performance scatter when measuring parts produced with the same target geometry. Particularly, it has become clear that the angle with which the punch is applied is a significant contributor to the scatter and thus must be focus of calibration.

Additionally, a non-destructive quality assurance of measuring the formed part can enable an improved prediction of the achieved performance of the specific part.

\section{Outlook}\label{sec:outlook}
A necessary next step is to validate the findings presented here experimentally beginning with surveying a significant number of formed parts by measuring their geometries. This data would allow the validation of the assumption that the indenter is mostly varying in angle and depth, which could be adapted if necessary while using the established simulation workflow. Furthermore, this data would allow adjusting the distribution of the geometric variances to be more realistic as they are currently assumed to be uniformly distributed. It would be of interest to consider other geometry-defining parameters as well. The process of cold forging creates various geometric imperfections that could also be considered, such as ridges. The resulting cross-sectional ellipses with surface irregularities can considerably influence wall thickness, and thus performance.

Subsequently a significant number of burst tests of measured samples allow the validation of the simulations at an individual part level. Additionally, this data allows to validate the correlation proposed between geometric variances and scatter in performance. Lastly, these tests would enable to verify the assumed correlation of maximum von Mises stress and burst failure and therefore ultimate performance.

Ultimately, the goal is to combine the findings proposed here with a model that incorporates the damage induced by the forming process itself. This would allow a more holistic approach when designing and dimensioning parts, potentially unlocking previously untapped lightweight potential.

\section*{Acknowledgements}
The authors gratefully acknowledge the funding by the Deutsche Forschungsgemeinschaft (DFG, German Research Foundation) in the framework of the Collaborative Research Centre CRC/TRR 188 ``Damage Controlled Forming Processes'' (project ID 278868966).

\subsection*{Data availability}

All relevant files can be found here:
\href{https://doi.org/10.5281/zenodo.17019105}{doi.org/10.5281/zenodo.17019105}

\nocite{*}
\printbibliography[heading=bibintoc,title={References}]

\end{document}